\documentstyle[12pt,epsfig]{article}
\def\beqra{\begin{eqnarray}} 
\def\eeqra{\end{eqnarray}}
\def\beqran{\begin{eqnarray*}} 
\def\eeqran{\end{eqnarray*}}
\def\beq{\begin{equation}}      
\def\eeq{\end{equation}}
\def\ds{\displaystyle}

\def\L{\Lambda}
\def\re#1{(\ref{#1})}
\def\D{\Delta}
\def\G{\Gamma}
\def\p{\partial}
\def\de{\delta}

\newcommand{\Tr}{\mathop{\mathrm{Tr}}}

\def\i{i}

\def\d{d}

\def\half{\mbox{\small $\frac{1}{2}$}}

\begin{document}
\sloppy

%
%
\begin{titlepage}
\begin{flushright}
CERN-TH/96-317\\
SHEP 96-31\\
\end{flushright}
\vspace{2cm}
\centerline{\Large\bf  Gauge-Invariant Renormalization Group}
\vspace{10pt}
\centerline{\Large\bf  at Finite Temperature}
\vspace{24pt}
\centerline{\large M. D'Attanasio}
\centerline{\it Department of Physics, University of 
Southampton,}
\centerline{\it Southampton SO17 1BJ, United Kingdom}
\vspace{20 pt}
\centerline{\large M. Pietroni}
\centerline{\it Theory Division, CERN}
\centerline{\it CH-1211 Geneva 23, Switzerland}

\vspace{24pt}
\begin{abstract}
We propose a gauge-invariant version of Wilson Renormalization 
Group for thermal field theories in real time.
The application to the computation of the thermal masses of the gauge bosons 
in an SU($N$) Yang-Mills theory is discussed. 
\end{abstract}
\vspace{6.3cm}
CERN-TH/96-317\\
November 1996
\end{titlepage}

\section{Introduction}
It is commonly believed that QCD at high temperature $T\gg\L_{\rm QCD}$
is in a deconfined phase, that is the system is composed by a gas of
quarks and gluons interacting with a coupling $\alpha_{\rm QCD}(T)\ll 1$.
Therefore perturbation theory is reliable at energy scales of order $T$.
This picture runs into problems when one probes the softer scale $gT$,
where contributions of the same order  are generated at every order
in perturbation theory \cite{Pisarski}.
At high temperature the relevant $gT$ contributions are obtained by resumming
the one-loop Feynman diagrams (Hard Thermal Loop resummation \cite{HTL}).

However, in non-Abelian gauge theories there is another major obstacle
in performing perturbative computations. Since the static magnetic fields
are not screened at HTL level, one finds wild infrared divergences at
order $g^2 T$, scale in which one expects such a magnetic mass to be 
dynamically generated by the interacting theory. Therefore there is a 
need for new resummation methods.

In ref.~\cite{noi} we proposed a possible approach.
The basic idea is to consider a ``cold'' $T=0$ quantum system and
add thermal fluctuations at lower and lower frequency scales, until
the non-perturbative scale $g^2T$ is reached. In this way the system
is thermalized at larger and larger distances until the scale $g^2T$ 
is reached ``smoothly''.
This corresponds to performing a coarse-graining \`a la Wilson \cite{Wilson}
for the thermal fluctuations. We will call such a method Thermal 
Renormalization Group (TRG).

This approach is inspired by continuum Wilson or ``Exact''
Renormalization Group (ERG) \cite{Polchinski}--\cite{Morris},
in which one starts with the microscopic ultraviolet action and adds
quantum modes with frequency larger than a coarse-graining scale $\L$.
By continuously varying $\L$ one gets the ERG flow equations in the form 
of partial differential equations in $\L$, which can be integrated to 
obtain the full quantum theory at $\L=0$.
The main difficulty associated to the application of ERG to $T=0$ gauge
theories is that it is not possible to introduce the infra-red 
cut-off $\L$ in a gauge-invariant way. Therefore the
Slavnov-Taylor identities of the theory are violated and one can only hope
to recover them once the cut-off is removed 
\cite{Becchi,Bonini2}--\cite{Reuter}.

One encounters similar difficulties when the ERG is applied to thermal
field theories. If one tries to perform a coarse-graining of both
quantum and thermal fluctuations, the ST identities are necessarily
violated by the coarse-graining scale $\L$ and it is much more difficult
to extract gauge-invariant quantities (see for instance
\cite{Reuter2}).

However, as explained above, the problem is simpler if one is interested in
resumming \`a la Wilson the thermal modes only, namely in TRG.
One assumes that the $T=0$ renormalized quantum field theory is known 
(from experiments, perturbation theory, lattice or other
approximation methods) and adds thermal fluctuations for 
frequencies $|\vec k|>\L$.
In this way one interpolates between the renormalized theory at $T=0$
($\L=\infty$, no thermal modes have been integrated out) and the 
theory in thermal equilibrium at temperature $T$ ($\L=0$, all
the thermal modes have been integrated out). 

For a finite value of the cut-off $\L$ one is describing a physical, 
{\it i.e.} renormalized, system out of thermal equilibrium, with a 
$\L$-dependent density matrix $\rho(\L)$.
{F}rom a physical point of view, the better status of gauge symmetry in this 
approach can be easily understood.
Indeed, the expectation values and the Green functions in an 
out-of-equilibrium system differ from those at $T=0$ (and from those in 
thermal equilibrium) because of the different weights of the 
{\it physical} states in the thermal average. 
But, as long as no unphysical states are introduced, the BRS 
variation and the thermal averaging are commuting operations and 
the Slavnov-Taylor identities are completely
analogous to the ones at $T=0$, provided the vacuum expectation values are 
interpreted as thermal averages with respect to the density matrix $\rho(\L)$
\cite{Chou}.

{F}rom a technical point of view, varying the density matrix corresponds to 
changing the temporal boundary 
condition that has to be given in order to invert  the quadratic part in 
the Lagrangian and define the free propagator. The Lagrangian itself is not 
modified and retains all its symmetries. 
This is to be contrasted with the usual 
formulation of the Wilson RG in which the introduction of the cut-off breaks
the BRS invariance of the Lagrangian explicitly
\cite{Becchi,Bonini2}--\cite{Reuter,Reuter2}. 

In order to implement this modification of the temporal boundary conditions 
we need a {\it real-time} formalism suited to describe a system with an 
arbitrary initial density matrix.
Therefore we find it appropriate to use the Closed Time Path (CTP) 
method \cite{Chou}--\cite{Boyanovsky}, 
which is a general formalism to deal with out of equilibrium problems.
The CTP technique is a generalization of the Real Time (RT) \cite{Niemi}
formulation of equilibrium thermal field theories and involves a doubling 
of the degrees of freedom.

In CTP the information about the initial state is encoded in source terms
concentrated at $t = t_0 \rightarrow -\infty$. 
By invoking an adiabatic switching-off for
the couplings, such information is contained as boundary conditions
in the free propagators.
By giving equilibrium boundary conditions, one recovers the
path-integral representation for the generating functional of the 
usual Real Time equilibrium thermal field theory given, for instance, in 
ref. \cite{Landsman}. 
We modify the boundary conditions by introducing a
momentum-dependent temperature $\beta(|\vec k|)$ such that
the path-integral representation for our coarse-grained
generating functionals can be obtained by substituting, in the equilibrium
path-integral, the Bose distribution for a cut-off one
\beq
N_b(|k_0|)\to N_b(|k_0|,\L)=\theta(|\vec k|-\L)\, N_b(|k_0|)\,.
\label{TRGBose}
\eeq
By varying with respect to $\L$ we will obtain the TRG flow equations
for the generating functional and for all the Green functions of the theory.

Integrating the flow equations from $\L = \infty$ 
\footnote{In practice, due to the 
Boltzmann suppression of high-frequency modes, an initial $\L \gg T$ will 
be required.} to $\L=0$, the resummed Green functions to all orders in 
thermal perturbation theory are obtained.

Another advantage of using the TRG in studies at high temperature
is that the boundary
condition for the flow equations at $\L \gg T$
is just the full quantum theory at $T=0$, which is directly related to the
physical parameters. The presence of the infra-red cut-off allows a reliable 
computation of the initial conditions in the framework of zero temperature 
perturbation theory.

A good and clear control over the initial conditions is
indeed a necessary requirement when one is interested  in small  thermal
effects such as the
generation of a magnetic mass for the gluon, which we expect to be $O(g^2 T)$.
On the other hand, using as initial condition the bare theory at some 
ultraviolet cut-off, as one is forced  to do in the ERG, the connection between
the physical parameters and the initial ones is  obtained only after 
the pure quantum corrections have been included via some approximation of the
flow equations at $T=0$.

There is another advantage in using the real time TRG instead of ERG
at finite temperature in the imaginary time. In a
non-interacting thermal gauge field theory it is not necessary to give
thermal fluctuations to the unphysical degrees of freedom, such as ghosts
and longitudinal gluons (only physical transverse particles are in contact
with the thermal bath and get heated). Therefore the Feynman rules to construct
perturbation theory can be simplified in such a way that only the propagator
of transverse gluons receives thermal boundary conditions. This is true
both in physical gauges (such as axial \cite{James1}--\cite{Nachbagauer})
and covariant gauges \cite{Landshoff}.
In this way only the transverse modes of the gluons have to be coarse-grained
and one gets simplified TRG flow equations. This approach is denoted
as TRG in physical phase space and is very useful if one wants to exploit
TRG as a real computational tool.

The paper is organized as follows. In Sect.~2 the CTP formulation for
an SU($N$) gauge theory is briefly discussed. In Sect.~3 the TRG flow 
equations are derived and the kernel is discussed.
In Sect. 4 we derive the ST identities and show that they are the same as those
for the theory without the cut-off on the thermal sector.
In Sect.~5 the computation of
the electric and magnetic screening masses at order $g^2T$ in the TRG 
formalism is discussed.
The appendices contain some technical details.

\section{Closed Time Path for pure gauge SU($N$) theory}
In this paper we will specialize to the case of an SU($N$) Yang-Mills
theory. The inclusion of fermions and other matter fields poses no
conceptual difficulties.

The action is
\beq
S=\int {\d}^4 x\,\biggr\{
-\frac 1 4 \left( F_{\mu\nu}  \cdot F_{\mu\nu}\right)
-\frac{1}{2\alpha}  {\cal F}^2
- \bar{c} \cdot {\cal M} \cdot c
\biggl\}\,,
\label{Scl}
\eeq
where
$$
F^a_{\mu\nu}(x)=\partial_\mu A^a_\nu -\partial_\nu A^a_\mu
+ g \left( A_\mu \wedge A_\nu\right)^a \,,
\;\;\;\;\;\;
\left(A_\mu \wedge A_\nu\right)^a=
\epsilon^{abc} A^b_\mu A^c_\nu \,,
$$
$$
F_{\mu\nu} \cdot F_{\mu\nu} =F_{\mu\nu} ^a F_{\mu\nu} ^a\,,
\;\;\;\;\;\;
D_\mu c =\partial_\mu c + gA_\mu \wedge c\,;
$$
${\cal F}[A_\mu]$ is the gauge-fixing term and ${\cal M}$ the Fadeev-Popov 
operator. The metric tensor $(1,-1,-1,-1)$ and the colour indices will
be understood in the following.

The action \re{Scl} is invariant under the BRS transformations 
\beq\label{BRS}
\delta A_\mu= \eta D_\mu c
\,,
\;\;\;\;\;\;\;
\delta c= -\eta \frac{g}{2} c \wedge c
\,,
\;\;\;\;\;\;\;
\delta \bar{c}= - \frac 1 \alpha \eta {\cal F}[A_\mu]
\,,
\eeq
with $\eta$ a Grassmann parameter.
In many formulae the fields and corresponding sources will be denoted by 
\beq
\Phi=( A,\, c,\, \bar c)\,,
\quad\quad
J=(j,\, \bar\chi,\, -\chi)\,.
\eeq
As we stated in the introduction, our formulation of the TRG is based on a 
deformation of the density matrix of the system implemented by introducing
an infra-red scale $\L$ such that for $\L \rightarrow \infty$ the system is at 
zero temperature 
and at $\L \rightarrow 0$ it is in thermal equilibrium at a temperature $T$.
For intermediate values of $\L$, only the high-momentum modes with 
$|\vec{k}|>\L$
are in thermal equilibrium, whereas the low-momentum ones are frozen at $T=0$.

In order to describe this system for a generic value of $\L$,
we shall use the CTP formalism introduced in \cite{noneq}, which is 
designed for a system prepared at some initial time $t_0$ 
with a generic density matrix $\rho$. In our applications the limit $t_0
\rightarrow -\infty$ is always 
understood.

The partition function can be represented as a path integral in which the 
time argument of the fields takes values on a path going from an initial
time $t_0$ to $t=+\infty$ along the real time axis, and then back to 
$t_0 - i\varepsilon$ infinitesimally below it. 
Indicating with $\Phi^1$ and 
$\Phi^2$ the fields with time arguments on the upper and lower pieces of the
time path, respectively, the partition function is given by
(for the proof of this and more details on the CTP method 
the reader is referred to \cite{Calzetta}): 
\beq
Z[J^1,J^2,\rho]=\int [{\d}\Phi]
\langle \Phi^1,t=t_0 | \rho|\Phi^2,t=t_0\rangle
\exp\i\left\{\half \Tr \Phi \cdot D_0^{-1} \cdot \Phi + 
S_{\rm int}[\Phi]+ \Tr J \cdot \Phi \right\}\,,
\label{ctp}
\eeq
where 
\[
[{\d}\Phi]=[{\d}\Phi^1][{\d}\Phi^2]\,,
\]
\[ 
\half\Tr\,\Phi \cdot D_0^{-1} \cdot \Phi = \sum_{i,j=1,2} 
\int \frac{{\d}^4 p}{(2 \pi)^4} \left\{ \half 
A_\mu^i(p)\, [{\bf G_0}^{-1}_{\mu\nu}(p)]^{ij} \,A_\nu^j(-p) 
+ \bar c^i(p)\, [{\bf S_0}^{-1}(p)]^{ij} \,c^j(-p) \right\}\,,
\]
\[
\Tr\,J\cdot\Phi = 
\int\frac{{\d}^4 p}{(2\pi)^4} \left\{
j_\mu^1(-p) A_\mu^1(p)+\bar\chi^1(-p)c^1(p)+\bar c^1(-p)\chi^1(p)
\:\:-\:\: 1\to 2
\right\}
\]
and $S_{\rm int}[\Phi] $ is the bare interaction action
\beq
S_{\rm int}[\Phi]=S_{\rm int}[\Phi^1]-S_{\rm int}[\Phi^2].
\eeq
${\bf G_0}_{\mu\nu}$ and ${\bf S_0}$ are the gauge and ghost free propagator, 
respectively (the $2\times 2$ CTP matrices are indicated in bold face).

In order to define the BRS transformation at the quantum level, 
source terms for the
composite operators appearing in \re{BRS}, that is
\beq
w_\mu^1 (D_\mu c)^1
-\frac{g}{2} v^1 c^1 \wedge c^1
\:\:-\:\: 1\to 2
\,,
\label{SBRS}
\eeq
have to be added to (\ref{ctp}).
In the following the dependence on the sources $w_\mu$ and
$v$ will be  understood.

In order to construct a perturbation theory for the path integral
\re{ctp}, one assumes that initial correlations at $t=t_0$ have a small
effect on the system at much larger times. This hypothesis is
equivalent to adiabatically switching off the interactions for 
$t\rightarrow t_0$ \cite{Landsman}. In this way 
the initial states $|\Phi^i,t=t_0\rangle$ ($i=1,\, 2$) are the {\it physical} 
states of the non-interacting theory (this requires special care in the
case of a gauge theory, where the Hilbert space is larger than the physical 
space---we will come back to this issue in the following). Moreover
the initial density matrix describes a free system and can then  be 
taken diagonal, that is
\beq
\rho\sim \exp \left\{ -\int{\d}^3 k \, \beta(\vec k) 
\sum_{{\rm phys.}\:{\rm states}} 
a^\dag_{\vec k} a_{\vec k} \right\}\,,
\label{rhoscalar}
\eeq
where as usual $a_{\vec k}$ is the annihilation operators for the 
free initial state $|\Phi(\vec k),t=t_0\rangle$ of 
momentum $\vec k$ (for simplicity we choose the same 
``temperature'' $\beta(\vec k)$ for all modes).
The density matrix \re{rhoscalar} is a
generalization of the equilibrium density matrix, which is obtained
for $\beta(\vec k)=\beta \omega_{|\vec k|}$.
The matrix elements of \re{rhoscalar} at $t=t_0$ can be represented in the
following way \cite{Calzetta}:
\beq\label{densmatrix}
\langle \Phi^1,t=t_0 |\rho|\Phi^2,t=t_0\rangle = \exp\left\{\i
\half\int K^{ij}\Phi^i\Phi^j \right\}\,.
\eeq
Therefore the statistical information on the initial state is all
contained in the non-local boundary source 
$K^{ij}(t_0, \vec{x}; t_0, \vec{x}^\prime)$, 
which is concentrated around $t=t_0$.
$K^{ij}$ is  just the boundary condition in time needed in order to invert
the d'Alembertian operator and to define the free propagator.
Notice that with a diagonal density matrix, as in \re{rhoscalar}, 
the boundary sources $K^{ij}$ can be computed exactly 
(see \cite{Calzetta} and below). 

Our goal is a path integral representation of a system with modes
thermalized for scales $|\vec k|>\L$. This can be obtained,
according to \re{TRGBose}, by choosing $\beta(\vec k)=\beta_\L(\vec k)$,
where
\beq
\beta_\L(\vec k)=\left\{
\begin{array}{cc}
\beta \omega_{|\vec k|} & {\rm for}\;\; |\vec{k}|>\L\\
&\\
+\infty & {\rm for}\;\; |\vec{k}|<\L
\end{array}
\right. \,.
\label{betaTRG}
\eeq
In this way the boundary source $K^{ij}$ becomes $\L$-dependent and
the partition function of such a system can be written as 
\beq
Z_\L[J^1,J^2,\rho]=\int [{\d}\Phi]
\exp\i\left\{\half \Tr \Phi \cdot D_{0,\L}^{-1} \cdot \Phi + 
S_{\rm int}[\Phi]+ \Tr J \cdot \Phi \right\}\,,
\label{ZL}
\eeq
where the appropriate source term $K_\L$ has been absorbed in 
\beq
D_{0,\L}^{-1} \equiv D_0^{-1} + K_\L \,.
\label{COP}
\eeq
In general, the free propagator \re{COP} can be computed by performing
the same manipulations as are done  in the well-known equilibrium case 
$\beta(\vec k)=$ constant \cite{Calzetta}. 
Therefore in a scalar theory \cite{noi} $D_\L$ is obtained from the 
one in the real time equilibrium 
formalism of Niemi and Semenoff \cite{Niemi}, with the obvious 
substitution $\beta\to\beta_\L(\vec k)$. This is equivalent to 
replacing the Bose-Einstein distribution function with the cut-off 
one as in \re{TRGBose}.\footnote{One could also
generalize the cut-off Bose distribution by using a smooth cut-off
function $\Theta(|\vec{k}|,\L)$ ($\Theta(|\vec{k}|, \L) \rightarrow 1$
for $|\vec{k}| \gg \L$ and $\Theta(|\vec{k}|, \L) \rightarrow 0$ for 
$|\vec{k}| \ll \L$). 
In the practical applications, however, we will use a Heavyside step 
function.}

When dealing with a gauge model, there is a further complication, namely,
after gauge fixing,
the Hilbert space and the set of physical states no longer coincide. As a 
consequence, the partition function cannot be generally defined as 
the trace over the whole Hilbert space of the density matrix $\rho$, 
but is instead given by
\beq
Z(\beta)=\Tr \left[{\rm P} \rho \right]\;,
\label{partition}
\eeq
where P is the projector onto the physical subspace.

We now consider separately the case of covariant and axial gauges.

\subsection{Covariant gauges}
In a covariant gauge, Fadeev-Popov (FP) ghosts and unphysical 
degrees of freedom populate the Hilbert space, and the projector P is 
non-trivial.
This problem is of course closely related to the question of which
temporal boundary conditions should be given to the FP ghosts and to the 
unphysical degrees of freedom of the gauge bosons, which could never
come into equilibrium with the thermal bath.

In equilibrium thermal field theory there are mainly two approaches to
handle this problem. The standard one (which we call the Bernard-Hata-Kugo
approach) consists in giving periodic boundary conditions to the
unphysical degrees of freedom \cite{Bernard, Hata}. The other one 
(which we call the 
Landshoff-Rebhan approach) consists in giving thermal boundary
conditions only to the physical (transverse) degrees of freedom 
\cite{Landshoff}.
We now describe how to generalize them to our TRG case.

(i) Bernard-Hata-Kugo approach.
The basic observation behind this approach is that 
the partition function can be rewritten as
\beq
Z(\beta)=\Tr \left[e^{-\pi Q_c} \rho \right] \;,
\label{pf}
\eeq
where $Q_c$ is the FP ghost charge. Moreover, the thermal average of 
any gauge-invariant operator ${\cal O}$,
\beq
\langle {\cal O} \rangle \equiv \Tr\left[{\rm P} \rho \,{\cal O}\right]/
 \Tr\left[{\rm P} \rho \right]
\label{ota}
\eeq
 is equal to
\beq
\langle {\cal O} \rangle = \Tr\left[ e^{-\pi Q_c} \rho \, {\cal O}\right]
/\Tr\left[e^{- \pi Q_c}\rho\right]\:\:\:\:\:\:\:\qquad{\rm if} 
\:\:\:\:\:\: [{\cal O}, Q_{\rm BRS}]=0\;,
\label{nta}
\eeq
where $Q_{\rm BRS}$ is the BRS charge.

Since any physical observable must be BRS-invariant, Hata and Kugo proposed
to modify the definition of a thermal average from eq. (\ref{ota}) to 
(\ref{nta}). This gives of course Green functions different from  those
obtained by using (\ref{ota}), but the result for physical quantities is 
unchanged. With this definition one can now derive path integral 
representations 
for the partition function and the Green functions by replacing $\rho$ with
$\rho \exp(-\pi Q_c)$ and  enlarging the set of states 
$|\Phi^i, t_0\rangle$ in (\ref{ctp}) to the full Hilbert space. 
Strictly speaking, Hata and Kugo considered the case of an equilibrium
density matrix $\rho$. However, as  is shown in appendix A, it is
straightforward to generalize their proof to the case in which $\rho$
has the form 
\beq
\rho\sim \exp \left\{ -\int{\d}^3 k \, \beta(\vec k) \left[
\sum_{\lambda=0}^3
a^{(\lambda)\dag}_{\vec k} a^{(\lambda)}_{\vec k}
+\half \bar{c}^{\dag}_{\vec k} c_{\vec k} 
-\half c^{\dag}_{\vec k} \bar{c}_{\vec k}
\right]\right\}\,,
\label{rhoHK}
\eeq
where $a^{(\lambda)}_{\vec k}$, $c_{\vec k}$ and 
$\bar{c}_{\vec k}$ are the annihilation operators for the free state of 
momentum $\vec k$ for gauge bosons with polarization index $\lambda$, 
ghost and antighost, respectively.
In the special case of thermal equilibrium, 
$\beta(\vec k)=\beta\omega_{|\vec k|}$,
this corresponds to assigning periodic boundary 
conditions to all the bosonic degrees of freedom and also to the FP 
ghosts.\footnote{This last prescription, formulated for the first 
time by Bernard \cite{Bernard}, comes from the fact that 
expression  (\ref{pf}) 
can be regarded as the partition function of a grand-canonical 
ensemble with purely imaginary chemical potential 
$\mu_{\rm FP}=\i \pi/\beta$ for FP ghosts \cite{Hata}.}
This is the approach that has been traditionally followed in the 
applications of perturbative thermal field theory to gauge theories.

As explained in the previous section the TRG formulation is obtained
by choosing the ``temperature'' $\beta(\vec k)$ as in \re{betaTRG}.
The computation of the boundary source $K_\L$ is completely analogous
to the equilibrium case. 
The free TRG gauge propagator is then the usual real time equilibrium 
gauge propagator \cite{Landsman}, with the Bose distribution replaced 
by the cut-off one in eq.~\re{TRGBose}, that is 
\beq
{\mathbf G_0}_{\L,\mu\nu}(k) = - 
\left(A_{\mu\nu}+B_{\mu\nu}+\alpha D_{\mu\nu}\right)\;
\biggl\{ {\mathbf Q}[\D_0(k)]
+ [\D_0(k) - \D_0^*(k)] \, N_b(|k_0|,\L)
\;{\mathbf B} \biggr\} \,,
\label{HK1}
\eeq
where we introduced the notations 
\beq
{\mathbf Q}[f(k)] = 
\left(
   \begin{array}{cc}
   f(k) &  (f(k) - f^*(k)) \theta(-k_0) \\
   &\\
   (f(k) - f^*(k)) \theta(k_0) & - f^*(k)
   \end{array}
\right) 
\label{nonthermal}
\eeq
\beq\label{B}
{\mathbf B}=
\left(
   \begin{array}{cc}
   1 & 1  \\
   &\\
   1 & 1
   \end{array}
\right) \,,
\eeq
and the free Feynman propagator
\[
\D_0=\frac{1}{k^2 + \i \varepsilon}\,.
\]
The usual basis of tensors
\beqra
&&
A_{\mu\nu}=g_{\mu\nu}-\frac{1}{(ku)^2-k^2}
[(ku)(u_\mu k_\nu+u_\nu k_\mu)-k_\mu k_\nu-k^2 u_\mu u_\nu]\,, 
\nonumber \\ &&
B_{\mu\nu}=\frac{- k^2}{(ku)^2-k^2} 
\left( u_\mu - \frac{(ku) k_\mu}{k^2} \right)
\left( u_\nu - \frac{(ku) k_\nu}{k^2} \right) \,,
\nonumber \\ &&
\label{ABCD} \\ &&
C_{\mu\nu}=\frac{1}{\sqrt{2[(ku)^2-k^2]}}
\left[\left( u_\mu - \frac{(ku) k_\mu}{k^2} \right) k_\nu +
\left( u_\nu - \frac{(ku) k_\nu}{k^2} \right) k_\mu \right] \,,
\nonumber \\ &&
D_{\mu\nu}=\frac{k_\mu k_\nu}{k^2}
\nonumber 
\eeqra
satisfy the multiplication properties
\beqra
&&A\cdot B=A \cdot C= A\cdot D=0\;, \;\;\;\;\; A+B+D=1\;, \nonumber \\
&&A\cdot A=A\;,\;\;\;B \cdot B=B\;,\;\;\;C\cdot C= \half (B+D)\;, \nonumber\\
&&\Tr C\cdot B= \Tr C\cdot D = 0 \;.
\label{table}
\eeqra
In the thermal reference frame we have $u_\mu=(1,0,0,0)$, and
therefore the only
non-vanishing components of $A$ are the spatial ones, giving the
transverse tensor
\beq\label{tensors}
A_{ij}=-\de_{ij}+\frac{k_i k_j}{|{\vec{k}}|^2}\,,
\eeq
In the $\varepsilon\rightarrow 0$ limit, we have
\beq
\D_0(k) - \D_0^*(k) \longrightarrow -2\i \pi \de(k^2)\,,
\label{deltafunct}
\eeq
that is, the thermal part of the tree-level propagator involves 
on-shell degrees of freedom only 
(only real particles belong to the thermal bath).
The free TRG ghost propagator is the same as for a scalar massless particle
(notice again the cut-off Bose distribution):
\beq
{\mathbf S_0}_\L(k) = - 
{\mathbf Q}[\D_0(k)]
- [\D_0(k) - \D_0^*(k)] \, N_b(|k_0|,\L)
\;{\mathbf B} \,,
\label{HK2}
\eeq

(ii) Landshoff-Rebhan approach.
A second possible formulation of thermal gauge theories in covariant gauges
is obtained by restricting the space of initial states
$|\Phi^i, t_0 \rangle$ to the physical one, as required by the original 
definition of the partition function, eq.~(\ref{partition}).
One has to recall that the initial state is given by {\it free} particles.
In the case of an SU($N$) gauge theory,
 this means that only the gauge boson
states with spatially transverse polarization will contribute to the
trace, and consequently only the transverse components of the gauge 
fields will couple to $K$-sources.
This corresponds to choosing an initial density matrix $\rho$
\beq
\rho\sim \exp\left\{-\int{\d}^3 k \,\beta(\vec k) 
\sum_{\lambda=1}^2
a^{(\lambda)\dag}_{\vec k} a^{(\lambda)}_{\vec k}
\right\}\,,
\label{rhoscalar2}
\eeq
where  the sum over $\lambda$ now runs  over transverse polarizations only.

In the case of thermal equilibrium, this choice leads to the 
Landshoff-Rebhan formulation of Real Time thermal field theory 
\cite{Landshoff}. 

The TRG formulation is again obtained by using the cut-off Bose
distribution. The TRG free propagator is then
\beq
{\mathbf G_0}_{\L,\mu\nu}(k) =  
-A_{\mu\nu}\;
{\mathbf P}_\L[\D_0(k)]
-\left(
B_{\mu\nu}+\alpha D_{\mu\nu}\right)\;
{\mathbf Q}[\D_0(k)]
\,,
\label{LR1}
\eeq
where we introduced the compact notation 
$${\mathbf P}_\L[\D_0]={\mathbf Q}[\D_0(k)]+[\D_0(k)-\D_0^*(k)]N_b(|k_0|,\L)
{\mathbf B}\;.$$

The free ghost propagator is simply the $T=0$ one:
\beq
{\mathbf S_0}(k) = - 
{\mathbf Q}[\D_0(k)]\,.
\label{LR2}
\eeq
Compared to the standard TRG propagators in (\ref{HK1}) and (\ref{HK2}), we
see that only the transverse component of the free gauge propagator
proportional to $A_{\mu\nu}$ gets a thermal 
structure, which incidentally is gauge-parameter-independent.

In the practical applications of the TRG in covariant gauges, 
the formulation \`{a} la Landshoff-Rebhan turns out to be more 
convenient, due to the absence of thermal ghost loops.

\subsection{Temporal Axial Gauge}
In physical gauges, such as  axial or Coulomb ones, the Hilbert space 
contains no
unphysical states, and then P=1. In particular the ghosts decouple.

The derivation of a path integral representation in these gauges then 
goes on analogously to what is done for the Landshoff-Rebhan case in
covariant gauges, with only the physical transverse gluon modes
included in the density matrix \re{rhoscalar}.

Here we are mainly interested in the Temporal Axial Gauge (TAG), which is 
obtained by taking ${\cal F}^a = u_\mu A^a_\mu$ in (\ref{Scl}), with 
$u_\mu=(1,0,0,0)$, and letting $\alpha\rightarrow \infty$.

The TAG is not a ``good'' gauge, in the sense that the gauge is not
completely fixed by the choice $A_0^a=0$. This gauge freedom manifests
itself in the spurious $1/k_0$ poles in the longitudinal part
of the free propagator. In the framework of 
perturbation theory, different prescriptions have
been proposed to circumvent these singularities. In particular in the
real time formalism at $T\neq 0$ the issue has been studied by James
in ref.~\cite{James1}.  As we will discuss in subsect.~4.2, when computing
the gauge-independent pole masses using the TRG, all the terms
containing these spurious poles cancel. So in our applications it is
not necessary to fix a particular prescription.

The free TAG TRG propagator in covariant form is given by \cite{James1}
\beq
{\mathbf G_0}_{\L,\mu\nu}=-A_{\mu\nu}\;{\mathbf P}_\L[\D_0(k)]
- {\bf {Q}}^\prime\left[H_{\mu\nu} \D_0(k)\right]\;,
\label{freepropTAG}
\eeq
where 
\beq
H_{\mu\nu} \equiv B_{\mu\nu} -\frac{\sqrt{2}|\vec{k}|}{k_0^\prime}
C_{\mu\nu}-\frac{|\vec{k}|^2}{k_0^{\prime 2}} D_{\mu\nu}\;,
\label{HMUNU}
\eeq
and ${\mathbf {Q}}^\prime$ and $k_0^\prime$ has been chosen for definiteness
according to James's
prescription, {\it i.e.}
\beq
{\mathbf Q^\prime}[f(k)] = 
\left(
   \begin{array}{cc}
   f(k) &  (f(k) - f^*(k)) \theta(-k_3) \\
   &\\
   (f(k) - f^*(k)) \theta(k_3) & - f^*(k)
   \end{array}
\right) 
\label{TAGmatrix}
\eeq
and $k_0^\prime = k_0 + i\varepsilon/k_3$.

\section{TRG evolution equations}
By taking the derivative with respect to $\L$ of \re{ZL} 
we obtain the evolution equation for $Z_\L[J]$,
\beqra
\L\frac{\p\:\:}{\p\L} Z_\L[J] &=& 
-\i (2\pi)^8 
{\rm Tr}\left\{ \half \L \frac{\p\:\:}{\p \L} D^{-1}_{0 \L} 
\cdot \frac{\de^2 Z_\L[J]}{\de J \de J} 
\right\} 
\label{evz}
\\
&=&-\i (2\pi)^8 {\rm Tr}\left\{
\half\frac{\de \:\:}{\de j_\mu} \cdot \L \frac{\p\:\:}{\p \L}
{\mathbf{G_0}}_{\L,\mu\nu}^{-1} \cdot\frac{\de \:\:}{\de j_\nu}  
Z_\L[J]\right\} \,.
\nonumber
\eeqra
In deriving the second  equation the ghost propagator has been assumed to be
$\L$-independent, as in TRG in physical phase space. Therefore all the
discussion below holds only in such a case.
However, the flow equation in the Bernard-Hata-Kugo approach can be 
obtained straightforwardly by including also the $\L$-dependence of 
the free  propagator of the ghost.

Choosing as initial conditions for $Z_\L[J]$ at 
$\L\gg T$ the full renormalized theory at zero 
temperature, the above evolution equation describes the effect of 
the inclusion of the thermal fluctuations at the momentum scale 
$|{\vec k}| = \L$. 
At a certain scale $\L$ we are 
describing a system in which only the high-frequency modes 
$|{\vec k}| >\L$ are in thermal equilibrium. 
The low-frequency modes do not feel the thermal bath and behave like 
zero-temperature, quantum modes.

We define as usual the cut-off effective action as the Legendre 
transform of the generating functional of the connected Green 
functions $W_\L[J]= -\i \log Z_\L[J]$: 
\beq
\half \Tr\,
\Phi \cdot ( {{D_0}}_\L^{-1} - {{D_0}}^{-1}) \cdot \Phi
+\G_\L[\Phi] = W_\L[J] - \Tr\,J\cdot\Phi \,,
\quad\quad
\Phi(p)=(2\pi)^4 \frac{\de W_\L[J]}{\de J(-p)}\,,
\label{action}
\eeq
where we have isolated the effect of the cut-off on the free part 
of the effective action 
and used for the classical fields the same notation as for the quantum 
fields.
\begin{figure}
\vspace{2 cm}
\mbox{\epsfig{file=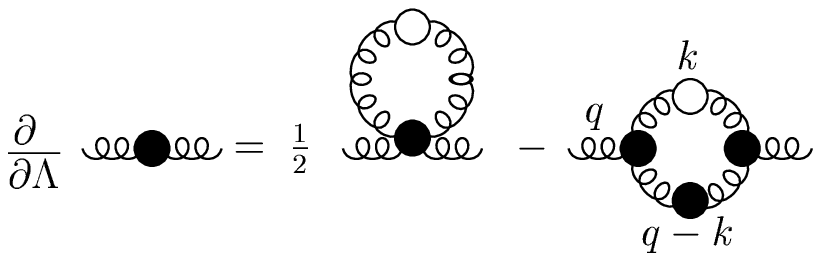,width=14cm}}
\vspace{1 cm}
\caption{\small{Flow equation for the self-energy in the physical
space. The black circles are the full 1-particle-irreducible vertices,
while the white circles represent the TRG kernel, given by 
eq.~\re{kerneltransv}.
}}
\label{se2}
\end{figure}
\newline
The evolution equation for the cut-off effective action $\G_\L[\Phi]$ 
can be derived straightforwardly from eq.~(\ref{evz}). This is done in
appendix B. The flow equations for the proper vertices are obtained  by 
derivation of the one for  the effective action. As an illustration, 
the flow equation for the gauge boson self-energy is
given in eq.~\re{phse} and
depicted in Fig.~\ref{se2}. We see that the recipe to obtain the TRG
flow equations is very
simple: one takes the one-loop graphs, substitutes every vertex and
propagator with the full ones, apart from a propagator that becomes
the RG kernel:
\beq
{\bf K}_{\L,\mu\nu} = \i
{\mathbf G}_{\L,\mu\rho} \cdot
\L\frac{\p\:\:}{\p\L} {\mathbf G_0}_{\L,\rho\tau}^{-1} \cdot 
{\mathbf G}_{\L,\tau\nu} \,,
\label{kerform}
\eeq
where the full propagator
\beq
{\mathbf G}_{\L,\tau\nu} = \left({\mathbf G_0}_{\L,\tau\nu}^{-1}-
{\mathbf\Sigma}_{\L,\tau\nu} \right)^{-1}
\label{fullpropgauge}
\eeq
appears.
Since the only $\L$-dependent part in ${\mathbf G}_{\L,\rho\tau}$ is 
proportional to the tensor $A_{\mu\nu}$ defined in eq. (\ref{ABCD}), and using 
the fact that  $A_{\mu\nu}$ is orthogonal to the remaining three tensors, 
the only component of the full propagator needed in order 
to compute the kernel is the one proportional to $A_{\mu \nu}$. It has the 
general form
\beq 
A_{\mu\rho} {\mathbf G}_{\L,\rho\nu}= 
-A_{\mu\nu} {\mathbf P}^\prime_\L[\D_{\rm T}(k)]\;,
\eeq
where we defined 
\beq
{\mathbf P}^\prime_\L[f(k)]={\mathbf Q}[f(k)]
+ [f(k) - f^*(k)] \, N_b^\prime(|k_0|, \L)
\;{\mathbf B} \,,
\label{P}
\eeq
and we introduced the ``longitudinal'' and ``transverse'' Feynman
propagators and self-energies
\beq
\D_{\rm L,T}(k)=\frac{1}{k^2-\Pi_{\rm L,T}(k)+\i\varepsilon}\,.
\label{LTse}
\eeq
The primes keep track of the fact that, in a generic gauge, the thermal 
structures appearing in the full propagator may correspond to an effective 
momentum-dependent temperature different from the physical one 
(see ref.~\cite{Landshoff2}). 
However, multiplying the three matrices appearing in (\ref{kerform}), it
is easy to see that all the terms proportional to $N_b^\prime(|k_0|, \L)$
drop, and the only temperature appearing in the kernel is the physical
one, coming from the free propagator. The kernel can then be written as
\beq
{\bf K}_{\L,\mu\nu}(k) =
- \i A_{\mu\nu} (\D_0 - \D_0^*)
\frac{\D_{\rm T} \D_{\rm T}^*}{\D_0 \D_0^*} \: 
\L \,\de(|{\vec k}| - \L) \,N_b(|k_0|)
\;{\bf B}.\label{k1} 
\eeq 
The above expression does not vanish in the $\varepsilon \rightarrow 0$ limit,
and we obtain
\beq
{\bf K}_{\L,\mu\nu}(k)=  -  A_{\mu\nu} \, \rho_{{\rm T},\L}(k) \epsilon(k_0)
\;\L \,\de(|{\vec k}| - \L) \,N_b(|k_0|) \;{\bf B} \label{kerneltransv}
\,,
\eeq
where $\epsilon(k_0)=\theta(k_0)-\theta(-k_0)$ and the transverse full
spectral function, 
\beq
i \rho_{{\rm T},\L}(k) = \epsilon(k_0) \, {\rm Disc} \, \D_{\rm T}(k),
\eeq
appears.
{F}rom (\ref{k1}) we note that the product of the kernel and the inverse of 
the full gluon 
propagator vanishes in the $\varepsilon \rightarrow 0$ limit. This fact will 
be used in the next section to derive ST identities in the present formalism.

We see that the kernel is exactly transverse and, in a covariant gauge,
 independent of the gauge-fixing 
parameter $\alpha$. The same form holds in the TAG.
It is also completely identical
(apart from the tensor $A$) to the kernel of a scalar particle, which
was studied in detail in ref.~\cite{noi}.

The physical meaning of this kernel is straightforward. As the 
infrared cut-off $\Lambda$ is lowered, 
the dispersion relation satisfied by the new modes coming into 
thermal equilibrium is not given by $k^2=m^2$ as in the $T=0$ free theory, 
but is determined by the full spectral function. In the approximation
in which one  neglects the imaginary part of the full self-energy this simply 
amounts to substitute the $T=0$ mass with the thermal, $\L$-dependent one.

In section 5 we will discuss the application of the 
formalism outlined in this paper to the computation of the electric
and magnetic mass in a high-temperature SU($N$) plasma.

\section{Slavnov-Taylor identities}
In general, the implementation of the Wilson RG in the framework of Quantum 
Field Theory requires special care when gauge theories are considered. 
The obvious reason is that the introduction of a momentum cut-off on  the 
quantum fluctuations immediately leads to a breaking of gauge invariance. 
The modified (inverse) propagator induces 
extra-terms into the ST identities, which recover their usual form
only in the $\L \rightarrow 0$ limit \cite{Bonini2, Ellwanger}.
To see this explicitely, we follow ref. \cite{Ellwanger} and perform a 
BRS-variation (\ref{BRS}) of the fields in the path integral of 
eq. (\ref{ZL}) in order to obtain the 
expression generating all the ST identities. 
For definitness, we discuss the case of general covariant gauges, but the 
discussion goes closely analogously for any class of gauges.
The ST identities for the proper vertices  are obtained by 
derivation of the expression
\beqra
&&\int \frac{d^4 p}{(2\pi)^4} 
\left\{\frac{\delta {\G}_\L}{\delta A_\mu^i(p)}
\frac{\delta {\G}_\L}{\delta w_\mu^i(-p)}
-\frac{\delta {\G}_\L}{\delta c^i(p)} 
\frac{\delta {\G}_\L}{\delta v^i(-p)} 
-\frac{i}{\alpha} p^\mu A_\mu^i(p)  \frac{\delta {\G}_\L}{\delta 
\overline{c}^i(-p)} \right.\nonumber\\
&&\label{STmamma}\\
&&\left.  - {\bf R}_{\mu\nu,\,ij}^\L(p) 
\frac{\delta^2 W_\L}{\delta w_\nu^i(-p) \delta j_\mu^j(p)}
-{\bf\tilde{R}}_{ij}^\L(p) \frac{\delta^2 W_\L}{\delta v^i(-p) \delta
\chi^j(p)} 
-\frac{i}{\alpha} p^\mu {\bf\tilde{R}}_{ij}^\L(p) 
\frac{\delta^2 W_\L}{\delta j_\mu^i(-p) \delta \overline{\chi}^j(p)} \right\} 
=0,
\nonumber
\eeqra
where we have defined ${\bf R}_{\mu\nu}^\L$ and ${\bf \tilde{R}}^\L(p)$ as the
differences between the inverses of the cut-off and uncut-off free
propagator for the gluons and the ghosts, respectively,
\beq
{\bf R}_{\mu\nu}^\L(p) \equiv {\bf G}_{\L, \mu\nu}^{-1}-
{\bf G}_{\mu\nu}^{-1}\,,\;\;\;\;\;\;\;\;
{\bf \tilde{R}}^\L(p) \equiv {\bf S}_{\L}^{-1}-
{\bf S}^{-1},
\eeq
where the color indices have been understood.

In the first line of eq.~(\ref{STmamma}) we recognize the usual expression for 
the ST identities obeyed by the full theory without the IR cut-off. The second 
line represents the breaking of BRS-invariance induced by the modification of 
the propagator. It vanishes for  $\L = 0$, as it should.

Now, the main point is that while  in the applications of the RG discussed
in refs.~\cite{Bonini2,Ellwanger,Reuter2}
the second line of eq.~(\ref{STmamma}) gives 
non-zero contributions for $\L \neq 0$, in the present 
approach  
it vanishes identically, 
and the cut-offed Green functions obey the same ST identities as
the ones of the full theory, the first line of eq.~(\ref{STmamma}). 
This can be understood by looking at the difference between the ${\bf R}$'s in
the two approaches. In refs.~\cite{Bonini2,Ellwanger,Reuter2} 
 the RG flow describes 
the effect of integrating out quantum fluctuations (or quantum plus thermal 
fluctuations in the case of ref.~\cite{Reuter2}, where the imaginary time 
formalism for thermal field theory has been considered). 
Therefore the cut-off is introduced
by modifying the free propagator according to 
\beq
G_{\mu\nu}(p) \rightarrow G_{\L,\mu\nu}(p) = G_{\mu\nu}(p) \Theta(p,\L),
\eeq
then 
\beq R_{\mu\nu}(p) =  G_{\mu\nu}^{-1}(p) (\Theta(p,\L)^{-1} - 1) ,
\eeq
and analogously for the ghost propagator.
As shown in ref.~\cite{Ellwanger} these expressions, inserted in the 
analogous of
eq.~(\ref{STmamma}) for the four-dimensional theory at zero temperature, 
give non-vanishing contributions such as a $\L$-dependent gluon mass. 

On the other hand in the present approach we are modifying only the thermal
part of the propagators (\ref{HK1}), (\ref{HK2}), or (\ref{LR1}),
by the substitution of eq.~(\ref{TRGBose}).
Considering for definitness the Landshoff-Rebhan approach, we have,
\beqra
{\bf R}_{\mu\nu}^\L &=& (N_b(|p_0|,\L)-N_b(|p_0|))  
\rho_{{\rm T},\L}(p) A_{\rho\sigma} \epsilon(p_0) \,
{\bf G}_{\L, \mu\rho}^{-1} \cdot {\bf B} \cdot
{\bf G}_{\L, \sigma\nu}^{-1}\nonumber\\
&&\\
&\equiv& (N_b(|p_0|,\L)-N_b(|p_0|)) \,{\bf G}_{\L, \mu\rho}^{-1} 
\cdot {\bf \tilde{k}}_{\L,\rho\sigma} \cdot
{\bf G}_{\L, \sigma\nu}^{-1}\nonumber \,,
\eeqra
where ${\bf \tilde{k}}_{\L,\rho\sigma}$ is proportional to the kernel
${\bf K}_{\L,\rho\sigma}$. 
{F}rom the discussion of the kernel in the previous paragraph we realize that 
the above expression vanishes in the $\varepsilon \rightarrow 0$ limit unless 
it is sandwiched between two full gluon propagators. 
This is not the case for  eq. (\ref{STmamma}) since, defining the amputated
field-dependent functional
${\bf \tilde{W}}_{\nu\rho}^{ik}$ by
\beq
\frac{\delta^2 W_\L}{\delta w_\nu^i(-p) \delta j_\mu^j(p)} \equiv 
{\bf \tilde{W}}_{\nu\rho}^{ik} {\bf G}_{\L,\rho\mu}^{kj}
\eeq
we have 
\beq
{\bf R}_{\mu\nu,\,ij}^\L(p) 
\frac{\delta^2 W_\L}{\delta w_\nu^i(-p) \delta j_\mu^j(p)} = 
(N_b(|p_0|,\L)-N_b(|p_0|)) \Tr  \left[{\bf \tilde{k}}_{\L,\rho,\sigma} \cdot 
{\bf G}_{\L, \sigma\nu}^{-1} \cdot {\bf \tilde{W}}_{\nu\rho} \right] \;,
\eeq
which vanishes in the $\varepsilon \rightarrow 0$ limit, since 
${\bf\tilde{W}}_{\nu\rho}^{ik}(p)$ does not contain a 
${\bf G}_{\L,\rho\mu}^{ik}(p)$ factor.
The same is true also in the Hata-Kugo approach, where the extra contributions
to the second line of eq.~(\ref{STmamma}) from longitudinal and gauge gluon
degrees of freedom, and from the ghosts, all vanish by he same reason 
explained above.

The Green functions defined in our approach are then gauge-invariant, in the
sense that they satisfy the usual ST identities, given by the first line 
of eq.~(\ref{STmamma}) even in the presence of a cut-off $\L\neq 0$. 

The fact that our TRG formalism respects ST identities for every 
value of $\L$ is a great advantage with respect to the other RG
methods, both from the conceptual point of view, since a physical 
meaning can be given to the TRG quantities for every $\L$ (one could 
also think of possible non-equilibrium applications), and from a
computational point of view, since the ST identities make it possible to
relate  propagator and vertices in a non-perturbative way.

\section{Application: the thermal masses}
The basic tool to study the behaviour of a high-temperature plasma 
is linear response theory \cite{LRT}. 
The perturbation of the thermal average of the vector 
potential induced by a weak source $\delta J^a_\mu$ is given by the 
well-known relation 
\beq
\delta \overline{A}_\mu^a(x) = \int dy \; \G_{\mu\nu}^{{\rm RET},ab}(x,y) 
\; \delta J^b_\nu(y)\,,
\label{Linear}
\eeq
where we further assume that $D^{ba}_\mu \delta J^a_\mu = 0$.

If we consider a source that is non-vanishing in only one direction in 
colour space, then the above relation is sufficient to determine the induced
perturbations on the field strength operator and consequently on the gauge 
invariant  chromo-electric and magnetic fields. 
The retarded propagator appearing in  (\ref{Linear}) is given by 
the difference between the 11 and 12 components of the full propagator 
\re{fullpropgauge}, which is the  object we will study. 
It is of course a gauge-dependent 
quantity, so the
question arises of what physical information can be derived from it.
As has been discussed in \cite{Rebhan, Kobes} the poles of the $A_{\mu\nu}$-
and $B_{\mu\nu}$- components of the full propagator are gauge-independent and
describe the exponential screening of chromo-electric and magnetic  fields,
respectively.

In a generic gauge, the full propagator can be written as
\beq
{\bf G}_{\L, \mu\nu} = {\mathbf a} A_{\mu\nu} + {\mathbf b} B_{\mu\nu}
+ {\mathbf c} C_{\mu\nu}+ {\mathbf d} D_{\mu\nu} \,,
\label{fullpropgen}
\eeq 
where ${\mathbf a}$, ${\mathbf b}$, ${\mathbf c}$ and ${\mathbf d}$ 
are gauge-dependent
$2\times 2$ matrices, the thermal structure of which is also
gauge-dependent \cite{Landshoff2}.  

Similarly,  the self-energy can be generally decomposed as
\beq
{\mathbf\Sigma}_{\L,\mu\nu}(k) = {\mathbf G_0}_{\L,\mu\nu}^{-1}(k) - 
{\mathbf G}_{\L,\mu\nu}^{-1}(k)\nonumber\\
= A_{\mu\nu} {\mathbf\Sigma}_{\rm T}(k)
+B_{\mu\nu} {\mathbf\Sigma}_{\rm L}(k)
+C_{\mu\nu} {\mathbf \Sigma}_c(k)
+D_{\mu\nu} {\mathbf \Sigma}_d(k)
\label{SD}
\eeq
(the $\L$-dependence of the 
${\mathbf\Sigma}$'s in the right-hand side has been omitted).
In general a Slavnov-Taylor identity will provide a relation between
the four matrices in \re{SD}, so that in a generic gauge the
self-energy will be determined by  three independent
functions. Nevertheless,
as discussed in \cite{Kobes} the pole structure of the temporal and
longitudinal modes is identical. Thus there are only two independent
physical modes; the spatially transverse and the longitudinal one.

The longitudinal and transverse self-energies are
\beqra
\ds {\mathbf\Sigma}_{\rm L}(k) &\equiv &
B^{\mu\nu}{\mathbf\Sigma}_{\L,\mu\nu} 
\,, \nonumber \\
\ds \label{LTSigma} \\
\ds {\mathbf\Sigma}_{\rm T}(k) &\equiv & \half 
A^{\mu\nu}{\mathbf\Sigma}_{\L,\mu\nu} \nonumber\,.
\eeqra
The real parts of the 11 components of ${\mathbf\Sigma}_{\rm L,T}$
coincide with the real parts of $\Pi_{\rm L,T}$ in eq.~\re{LTse}, 
which will denoted as $\Pi_{\rm L,T}^R$.

The two physical poles of the full propagator are then identified by the 
equations
\beq
k^2 - \Pi_{\rm L}^R(k) = 0 \,,
\qquad\qquad
k^2 - \Pi_{\rm T}^R(k) = 0 \,.
\label{poleq}
\eeq
We are interested in the screening of static chromo-electric and magnetic
 fields. The relevant poles are then given by
\beqra
m_{\rm L}^2(\L) &=& \left.\Pi_{\rm L}^R(q)
\right|_{q=q_{\rm L}}\,,
\nonumber \\
&&\label{poles}\\
m_{\rm T}^2(\L) &=&  \left.\Pi_{\rm T}^R(q)
\right|_{q=q_{\rm T}}\nonumber\;,
\eeqra
where $q_{\rm L,T}=(q^0_{\rm L,T}=0,\,|\vec{q}_{\rm L,T}|^2 =
-m^2_{\rm L,T})$.

The masses defined in \re{poles} are gauge-fixing-independent 
for any value of $\L$, as can be shown by strictly
following  ref.~\cite{KKR},
and considering variations in the gauge-fixing term which vanish for
$t \rightarrow t_0$.

The evolution equations for $m_{\rm L}^2$ and $m_{\rm T}^2$ can then
be written down as
\beqra
\ds \L\frac{\partial\;\;\;}{\partial \L} m_{\rm L}^2 &=& \ds 
\left. \frac{B^{\mu\nu}(q)
F_{\mu\nu}(q)}{1+\frac{\partial \Pi_{\rm L}^R(q)}{\partial
|\vec{q}|^2}}
\right|_{q=q_{\rm L}}\,,
\nonumber\\
&&\label{eveqm}\\
\ds \L\frac{\partial\;\;\;}{\partial \L} m_{\rm T}^2 &=& \ds
\left. \frac{\half A^{\mu\nu}(q)
F_{\mu\nu}(q)}{1+\frac{\partial \Pi_{\rm T}^R(q)}{\partial
|\vec{q}|^2}}
\right|_{q=q_{\rm T}} \;,
\nonumber
\eeqra
where (see eqs.~\re{phse} and \re{Gse})
\beqra
&&F_{\mu\nu}(q) \equiv 
\L\frac{\partial\;\;\;}{\partial \L} 
{\mathbf \Sigma}_{\L,\mu\nu}^{11}(q)=
\half \int \frac{d^4k}{(2\pi)^4}  
{\bf K}_{\L,\rho\sigma}(k) \left[\G^{11ij}_{\L,\mu\nu\rho\sigma}(q, -q,
k, -k) \right.\nonumber\\
&&\qquad\qquad\left. - 2 \G^{1ik}_{\L,\mu\rho\lambda}(q, -k, k-q) 
{\mathbf G}_{\L,\lambda\tau}^{kl}(q-k)  
\G^{1jl}_{\L,\nu\sigma\tau}(-q, k, q-k)\right]\,.
\label{FMUNU}
\eeqra
To simplify the notation, the colour structure has been understood.
In appendix C we verify explicitly that the running does not introduce any 
gauge dependence. Namely, we show that, when contracted with $B^{\mu\nu}(q)$
($A^{\mu\nu}(q)$) and computed at $q=q_{\rm L}$ ($q_{\rm T}$), 
$F_{\mu\nu}(q)$ gives a gauge-independent result. 

The system (\ref{eveqm}) has to be integrated from some initial 
$\L_0 \gg T$ down to $\L = 0$.
At $\L_0$ we have to insert the zero-temperature propagators and vertices 
in the right-hand-sides. These can be reliably computed using perturbation
theory, since,  
due to the momentum delta-function in the kernel, eq. (\ref{kerneltransv}), 
the loop momentum  is forced to be $|\vec{k}|=\L_0\gg T\gg\L_{\rm QCD}$ (the 
external momentum is $|\vec{q}|=m_{{\rm L},{\rm T}}(\L_0) \ll \L_0)$. 
In other words, the 
cut-off serves as IR regulator for the $T=0$ theory.
As we decrease $\L$ at $O(g T)$, an IR cut-off is generated in the 
longitudinal sector (the longitudinal, or Debye,  mass), 
whereas in the transverse sector a magnetic mass of $O(g^2 T)$ is believed to 
emerge non-perturbatively. 

The resummation of the thermal fluctuations from $\L\simeq T$ to 
$\L \simeq gT$ corresponds to the resummation of  Hard Thermal Loops
performed by Braaten and Pisarski \cite{HTL}. To go beyond that approximation,
 thermal fluctuations from $gT$ to $g^2 T$ have to be integrated out. 
In a perturbative framework, like HTL resummation, the different scales are
not clearly separated, and this is reflected, for instance, in the 
pathologies (gauge-dependence and logaritmic divergence) of the next-to-leading
correction to $m_{\rm L}$ \cite{Rebhan}, which require the ad-hoc introduction
of a magnetic mass term.
In the next
subsections we outline how this non-perturbative problem can be tackled
in the TRG formalism, both in covariant and physical gauges.

\subsection{Covariant gauges}
The class of covariant gauges is specified by taking 
${\cal F}^a= \partial^\mu A^a_\mu$ in eq.~(\ref{Scl}). 
The four matrices in  (\ref{fullpropgen}) are then
restricted by the ST identity:
\beq
k^\mu k^\nu G_{\L,\mu\nu} = \alpha\;.
\eeq
In Landau gauge ($\alpha=0$) we have the stronger relation 
$k^\mu  G_{\L,\mu\nu}=0$.

In order to obtain an explicit form for the full propagator to be inserted 
in the evolution equations for the thermal masses, we will neglect the
imaginary part of the longitudinal gluon self-energy, which allows us
to write the full propagator as 
\beq
{\mathbf\D}^0_{\L,\mu\nu}(k) = 
A_{\mu\nu} \; {\mathbf P}_\L[\D_{\rm T}] 
+ B_{\mu\nu} \; 
{\mathbf Q}[\D_{\rm L} + \alpha c^2 \D_{\rm L}^2 \D_0] 
- \alpha  C_{\mu\nu} \; {\mathbf Q}[c \D_{\rm L} \D_0] + 
\alpha D_{\mu\nu} \; {\mathbf Q}[\D_0] \,,
\label{fullprophoton}
\eeq
where $c(k)$ is a gauge-dependent function.

Due to the presence of the full three  and four gluon vertices (see
Fig.~1), and of the unknown function $c$, the system of coupled 
differential equations \re{eveqm} is not closed. The knowledge 
of the $\L$-dependent vertex functions would of course require the 
solution of their evolution equations which, in turn, involve 
higher vertex functions.

In order to get a manageable approximation, the system has to be truncated at
some level, in practice by using a suitable approximation of the three and 
four gluon vertices. 
A crucial requirement on this approximation is that it respects the 
Slavnov-Taylor equation for the full vertices to the required level of 
accuracy. This is essential if one wants a gauge-independent
determination of $m_{\rm L,T}^2$ which are
gauge-independent quantities in the full theory.
In practice we are interested in computing the squared masses at $O(g^4T^2)$. 
Looking at eq. (\ref{eveqm}) we see that the most dangerous $\alpha$-dependent
contribution comes from the $D_{\mu\nu}$-term in the propagator of eq.
(\ref{fullprophoton}). If we just use the tree-level 
expression for the trilinear vertex, then, since 
\beq
(k-p)_{\lambda} \G_{\mu\rho\lambda}^{\rm tree}(p,-k,k-p)
= g_{\mu\rho}(k^2-p^2)+ p_\mu p_\rho- k_\mu k_\rho 
\eeq
(the group and RT indices of this vertex are trivial),
it is easy to see that this $\alpha$-dependent term gives  a
contribution $O(m_{\rm T}^2)$ to $F_{\mu\nu}(q_{\rm T})$, 
{\it i.e} of the same order as the quantity we want to compute. 

A way to improve the unsatisfactory approximation of tree-level vertices
would be to enlarge the system of differential equations to include those for 
$\G_{\L,\mu\nu\rho}$ and $\G_{\L,\mu\nu\rho\sigma}$, 
setting to zero all the higher-order 
vertices appearing in the new equations.
In the next subsection we will see that, thanks to the simplicity of the 
ST identities, the TAG offers a practically
more appealing alternative.

\subsection{Temporal Axial Gauge}
The ST identities for the three and four gluon vertices in the TAG are
\beqra
&&
p_\lambda \left[ \G_{\L,\lambda\mu\nu}^{abc}(p,q,r)
\right]^{1ij}=-\i f^{abc} \left\{
\left[\tilde{\bf G}^{-1}_{\L,\mu\nu}(q)\right]^{1i} \delta^{1j}
-\left[\tilde{\bf G}^{-1}_{\L,\mu\nu}(r)\right]^{1j} \delta^{1i}
\right\}\,,
\label{TAGST1}
\\
&&
p_\mu \left[ \G_{\L,\mu\nu\rho\sigma}^{abcd}(p,q,r,s)
\right]^{1jkl}
=\i f^{ade} \left[ \G_{\L,\sigma\nu\rho}^{ebc}(p+s,q,r)\right]^{1jk}\delta^{1l}
\nonumber \\ && \qquad
+\i f^{ace} \left[ \G_{\L,\rho\nu\sigma}^{ebd}(p+r,q,s)\right]^{1jl}\delta^{1k}
+\i f^{abe} \left[
\G_{\L,\nu\rho\sigma}^{ecd}(p+q,r,s)\right]^{1kl}\delta^{1j}
\;.
\label{TAGST2}
\eeqra
In \re{TAGST1} we have defined $\tilde{\bf G}^{-1}_{\L,\mu\nu}(q)
\equiv {\bf G}^{-1}_{\L,\mu\nu}(q)-u_\mu u_\nu/\alpha\;{\bf Q}[1]$, 
which is well-behaved in the TAG limit $\alpha \rightarrow 0$.

Neglecting again the imaginary part of the longitudinal self-energy, 
the full propagator can be written as
\beq
{\mathbf G}_{\L,\mu\nu}=-A_{\mu\nu}\;{\mathbf P}_\L[\D_{\rm T}(k)]
-{\bf Q}^\prime\left[H_{\mu\nu} \D_{\rm L}(k)\right] 
\,,
\label{fullpropTAG}
\eeq
where $H_{\mu\nu}$ has been defined in \re{HMUNU}.

The self-energy is given by
\beq
{\mathbf \Sigma}_{\L, \mu\nu}(k) = A_{\mu\nu} \,
{\mathbf P}^{-1}_\L[\D_{\rm T}(k)^{-1} - k^2] + 
B_{\mu\nu}\,{\bf Q^\prime}^{-1}\left[ \D_{\rm L}(k)^{-1}-k^2 \right] \,.
\eeq
Now, as promised in subsect.~3.2, we show that, as a consequence of the
ST identity \re{TAGST1}, the
spurious poles in $1/k_0^\prime$ and $1/(k_0^\prime)^2$ appearing in
$H_{\mu\nu}$ do not give any contribution to the running of the two
physical masses $m_{\rm L}^2$ and $m_{\rm T}^2$. 
Indeed, from \re{HMUNU} we see
that these poles appear as coefficients of the tensors $C_{\mu\nu}(k)$
or $D_{\mu\nu}(k)$, and then they are always multiplied by $k_\mu$ or
$k_\nu$. So the analogous singularities appearing in the propagator 
${\mathbf G}_{\L,\lambda\tau}(q-k)$ in eq.~\re{FMUNU}
are proportional to $(q-k)_\lambda$ or $(q-k)_\tau$. 
Applying the ST identity \re{TAGST1}, at least one of the two full
three-gluon
vertices in \re{eveqm} can be written as the difference of the 
inverse propagator $\tilde{\bf G}^{-1}$ at the loop momentum $k$ and the
one at the external momentum $q$. 
The first one vanishes when multiplied by the
kernel ${\bf K}_{\L,\rho\sigma}(k)$. The second one vanishes at 
$q=q_{\rm L}$ if $F_{\mu\nu}(q)$ is contracted with $B_{\mu\nu}(q)$ 
and at $q=q_{\rm T}$ 
if it is contracted with $\half A_{\mu\nu}(q)$, so it does not contribute
to the evolution equations for $m_{\rm L}^2$ and $m_{\rm T}^2$.
The cancellation of all spurious singularities can also be 
proved for the denominators in \re{eveqm} in a closely analogous way. 

In the past, the simple form of the ST identities \re{TAGST1} and 
\re{TAGST2} has motivated the choice of the TAG in non-perturbative 
studies of the infrared limit of the gluon propagator, both at $T=0$ 
\cite{Baker} and at $T\neq 0$ \cite{Kajantie,Kalashnikov}. 
These studies were based on truncations of the exact Schwinger-Dyson
equations for the self-energy. In \cite{Baker} and \cite{Kalashnikov} 
the observation was made that the transverse part of the full 
three-gluon vertex, defined by
\beq
p_\lambda \G_{\L,\lambda\mu\nu}^{({\rm T})abc}(p,q,r)=0 ,
\label{3gT}
\eeq
and the requirement that $ \G_{\L,\lambda\mu\nu}^{({\rm T})abc}$ be free of 
kinematical singularities, vanishes when any of the external momenta
goes to zero. Then, in the infrared regime  of interest here,
the relevant part of the three-gluon vertex is given by the
longitudinal one, which is completely determined in terms of the
inverse full gluon propagator by means of eq.~\re{TAGST1} \cite{Kim}.
By expressing the four gluon vertex in terms of the longitudinal
three-gluon one, the set of infinitely many Schwinger-Dyson equations is
turned into two coupled equations for $m^2_{\rm L}$ and $m^2_{\rm T}$. 

In the present framework the same approximation of the $\L$-dependent 
three  and four gluon vertex can be implemented into the exact
(non-truncated) equations \re{eveqm}. Since the vertices are fully
determined by $\Pi_{\rm L}(k)$ and $\Pi_{\rm T}(k)$, the possibility 
arises of closing the system of
differential equations for the thermal masses. The simplest
approximation would be to take $\Pi_{\rm L,T}(k)\simeq m^2_{\rm L,T}$, in
which case one is left with only two coupled ordinary differential equations
for $m_{\rm L}^2$ and $m_{\rm T}^2$. 

This is a good starting point for a non-perturbative computation of the
screening masses. In a forthcoming paper  we will discuss this
and more refined approximations and present the results of a numerical
computation.

\section{Conclusions}
In this paper we have discussed the extension to gauge theories of the 
Wilson RG-inspired formalism introduced in ref.~\cite{noi}. The main 
attractive feature of this formalism is given by the fact that
introducing the infra-red cut-off on the thermal sector of the theory
does not
break gauge invariance. Namely, even in the presence of a non-zero
thermal IR cut-off $\L$, the ST identities are formally the same as 
at $T=0$, provided the Green functions are interpreted as thermal 
averages with respect to a density matrix $\rho(\L)$. 

This provides the basis for a gauge-invariant renormalization flow, in which
the running does not introduce extra gauge dependence. In computing a gauge 
dependent quantity, the result at the end of the  running ($\L=0$) will be 
the finite temperature counterpart of the same quantity computed at $T=0$ in
the same gauge. On the other hand, the TRG running for a properly defined gauge
independent quantity will be completely gauge independent.

{F}rom a practical point of view, a gauge invariant RG flow presents the 
great advantage  that ST identities can be used as a guide to 
reduce the number of independent invariants  to be included in approximation 
schemes based on truncations of the exact evolution equations.

The need for a proper control of gauge invariance in computation methods for 
the thermal  masses in hot QCD has recently been emphasized by many authors 
(see for instance \cite{Jackiw}).
We have discussed this problem in our formalism, pointing out that an
approximation of the full set of TRG equations based on lowest-order
truncations, in which only the running of the two-point functions is included,
leads to a gauge-dependent effect of the same order as the subleading 
corrections we are interested in. 
Due to the simplicity of the ST identities, the TAG turns out
to be a more appealing gauge, in which improved vertices can be defined in 
terms of inverse propagators and a closed set of approximate equations can be 
obtained. 
Remarkably, the HTL propagators and vertices are gauge independent and satisfy
the same ST identities as those of the TAG \cite{HTL}. Then, in stepping
 beyond HTL rsummation, one could  impose that the same identities 
are satisfied also by the new propagator and vertices, even in 
covariant gauges.

Work along these lines is now in progress and will be presented in a 
forthcoming paper.

\noindent
{\bf Acknowledgements}

We thank A. Hebecker and T.R. Morris for useful discussions.
M. D'A. would like to thank PPARC for financial support and
F. Zwirner and CERN for kind hospitality.
M.P. would like to thank Southampton Univ. for kind hospitality and
acknowledges support from the EC ``Human Capital and Mobility'' 
programme.

\appendix

\section{TRG generalization of Hata-Kugo proof}
The partition function \re{partition} can be rewritten as 
\beq
Z(\beta)=\Tr[{\rm P} e^{-\pi Q_c} \rho]\,,
\label{Zapp}
\eeq
since P projects onto the subspace with FP number $0$. Then we notice
\cite{Hata} that P can be written as 
\[
{\rm P}={\rm 1}-\left\{ Q_{\rm BRS},\:{\rm R} \right\}\,,
\]
where R is a suitable operator of FP number $-1$.
By inserting this in \re{Zapp} we find that the contribution
containing R vanishes
\beqran
&&
\Tr[\{ Q_{\rm BRS},\:{\rm R} \}\, e^{-\pi Q_c} \rho]
\\ && \quad 
=\Tr[{\rm R} \, e^{-\pi Q_c} \rho \, Q_{\rm BRS}]+
\Tr[{\rm R} \, Q_{\rm BRS} \, e^{-\pi Q_c} \rho] \qquad\qquad\qquad
{\rm (cyclicity)}
\\ && \quad 
=\Tr[{\rm R} \, e^{-\pi Q_c} \, Q_{\rm BRS} \, \rho]+
\Tr[{\rm R} \, Q_{\rm BRS} \, e^{-\pi Q_c} \rho] \qquad\qquad\qquad
{\rm (since} \:[Q_{\rm BRS},\:\rho]=0)
\\ && \quad 
=\Tr[{\rm R} \, \{e^{-\pi Q_c},\: Q_{\rm BRS}\} \, \rho]=0
\qquad\qquad\qquad\qquad\qquad\quad
{\rm (since} \:[Q_{\rm BRS},\:Q_c]=\i Q_{\rm BRS})
\eeqran
The crucial point is that the non-equilibrium density matrix $\rho$
commutes with the BRS charge, due to the adiabatic switching-off of
the interaction at $t=t_0$.

Therefore one finds \re{pf}.

\section{TRG equation for the cut-off effective action}
In this appendix we derive the TRG flow equations for a general gauge.

By deriving eq.~\re{action} with respect to $\L$ and using \re{evz} 
we get
\beq\label{evact1}
\L\frac{\p\:\:}{\p\L} \G_\L[\Phi]=
-\i(2\pi)^8 
{\rm Tr}\left\{ \half \L \frac{\p\:\:}{\p \L} {\bf G_0}^{-1}_{\L,\mu\nu} 
\cdot \frac{\de^2 W_\L[J]}{\de j_\mu \de j_\nu} 
\right\} \,.
\eeq
In the RHS of this equation there is a field-independent piece
coming from the full propagator. We isolate it and define a new
functional $\bar{\bf\G}$:
\beq
(2\pi)^8
\frac{\de^2 W_\L[J]}{\de j_\mu \de j_\nu} 
=(2\pi)^4 \,{\bf G}_{\L,\mu\nu} - 
{\bf G}_{\L,\mu\rho}\cdot
\bar{\bf\G}_{\L,\rho\sigma}\cdot{\bf G}_{\L,\sigma\nu}\,, 
\label{Gammabar}
\eeq
where ${\bf\Delta}$ is the full propagator \re{fullpropgauge}.
In this way the evolution equation \re{evact1} becomes (dropping
the field-independent term)
\beq
\L\frac{\p\:\:}{\p\L} \G_\L[\Phi]=\half
{\rm Tr}\left\{{\bf K}_{\L,\mu\nu}\cdot\bar{\bf\G}_{\L,\mu\nu}\right\}\,,
\label{evact2}
\eeq
where the kernel ${\bf K}$ is given in \re{kerform}.

We now have to express the functional in the RHS of \re{evact2}
in terms of $\Phi$. From the definition \re{action} one has
\beq
-j_\mu={\bf G_0}^{-1}_{\L,\mu\nu} \cdot A_\nu
+(2\pi)^4 \frac{\de\G_\L}{\de A_\nu}\,.
\eeq
Deriving this relation with respect to $j_\nu$, it follows
\beq
(2\pi)^4 
\frac{\de^2  W_\L}{\de j_\mu \de j_\nu}
\left({\bf G_0}^{-1}_{\L,\nu\rho}+(2\pi)^4 
\frac{\de^2\G_\L}{\de A_\nu \de A_\rho}\right)
=g_{\mu\rho}\,.
\label{inverse}
\eeq
We now isolate the self-energy
\beq
(2\pi)^8 \frac{\de^2 \G_\L[\Phi]}{\de A_\mu \de A_\nu}=
(2\pi)^4 {\bf \Sigma}_{\L,\mu\nu}+
{\bf\G}_{\L,\mu\nu}^{\rm int}[\Phi]\,. \label{Gammaint}
\eeq
Substituting eqs.~\re{Gammabar} and \re{Gammaint} into \re{inverse},
one finds
\beq
\bar{\bf\G}_{\L,\mu\nu}[\Phi]={\bf\G}_{\L,\mu\nu}^{\rm int}[\Phi]
-{\bf\G}_{\L,\mu\rho}^{\rm int}[\Phi]\cdot
{\bf G}_{\L,\rho\sigma}\cdot\bar{\bf\G}_{\L,\sigma\nu}[\Phi]
\,.\label{Gbar1}
\eeq
The evolution equations for the various vertices can be found by
expanding in powers of $\Phi$. 
As an example we give the flow equation for the gauge boson
self-energy (see Fig.~\ref{se2}).
By deriving \re{evact2} with respect to $A^1_\mu(p)$, $A^1_\nu(-p)$, and
setting the fields to zero, we get
(in the following the gauge group indices $a$, $b$, $c$, $d$ are
explicit, and no summation is made on the index $a$) :
\beq
\L\frac{\p\:\:}{\p\L} \left[ {\bf\Sigma}^{aa}_{\L,\mu\nu}(p)\right]^{11}= 
 \int\frac{{\d}^4 k}{(2\pi)^4} 
\half {\bf K}_{\L,\rho\sigma}^{ij}(k) \cdot
\left.\frac{\de^2\overline{\bf\G}^{ij}_{\L,\rho\sigma}}{\de 
A_\mu^{a,1}(p)\de A_\nu^{a,1}(-p)}\right|_{\Phi=0}
\,,
\label{phse}
\eeq
where the vertex of $\bar{\bf\G}$ is found from \re{Gbar1} 
\beqra
&&
\left.\frac{\de^2\overline{\bf\G}^{ij}_{\L,\rho\sigma}}{\de A_\mu^{a,1}(p)
\de A_\nu^{a,1}(-p)}\right|_{\Phi=0}= 
\G_{\L,\mu\nu\rho\sigma}^{aabb,11ij}(p,-p,-k,k)
\nonumber \\ 
&& \qquad
-2\,
\G_{\L,\mu\rho\lambda}^{abc,1ik}(p,-k,k-p)
\left[{\bf G}_{\L,\lambda\tau}\right]^{kl}(p-k)
\G_{\L,\nu\sigma\tau}^{abc,1jl}(-p,k,p-k)
\,,
\label{Gse}
\eeqra
where we have the three and four gluon vertices 
\[
\G_{\L,\mu\nu\rho}^{abc,ijk}(p,q,r) =
\left.\frac{\de^3\G_\L[\Phi]}
{\de A^k_\rho(r)\de A_\nu^j(q)\de A_\mu^i(p)}
\right|_{\Phi=0}\,,
\]
\[
\G_{\L,\mu\nu\rho\sigma}^{abcd,ijkl}(p,q,r,s) =
\left.\frac{\de^4\G_\L[\Phi]}
{\de A^l_\sigma(s)\de A^k_\rho(r)\de A_\nu^j(q)\de A_\mu^i(p)}
\right|_{\Phi=0}\,.
\]

\section{Gauge-fixing independence of the TRG equations for $m_{\rm L}^2$ 
and $m_{\rm T}^2$}
We are interested in the effect of a change of gauge-fixing function:
\beq
{\cal F} \rightarrow {\cal F} + \Delta {\cal F}\, ,
\label{gaugefix}
\eeq 
on the evolution equation for the self-energy.

Following ref.~\cite{KKR} we will start from the basic relation
\beq
\Delta W^{\cal F}[J] = - J_i  \Delta \chi_i\,,
\label{deltaW}
\eeq 
where $\Delta W^{\cal F}[J]$ is the change in the generating
functional
$W_\L[J]$ under the transformation (\ref{gaugefix}), and
\beq
\Delta \chi_i = \langle D_i \cdot {\cal M}^{-1} \cdot \Delta {\cal F}
\rangle \,.
\label{deltachi}
\eeq

Under the same change (\ref{gaugefix}) the evolution equation for the 
effective action $\G_\L[\Phi]$ is modified by the extra terms
\beq\label{change}
\L\frac{\p\:\:}{\p\L} \Delta\G_\L[\Phi] = H^1_\L[\Phi] + H^2_\L[\Phi]
\eeq
where
\beqra
H^1_\L[\Phi]&=&
-\i(2\pi)^8 
{\rm Tr}\left\{ \half \L \frac{\p\:\:}{\p \L} {\bf G_0}^{-1}_{\L, ij} 
\cdot \Delta\frac{\de^2 W_\L[J]}{\de j_i \de j_j}\right\} \label{changegamma}\\
H^2_\L[\Phi]&=& 
-\i(2\pi)^8 
{\rm Tr}\left\{ \half \L \frac{\p\:\:}{\p \L} \Delta 
{\bf G_0}^{-1}_{\L, ij} 
\cdot \frac{\de^2 W_\L[J]}{\de j_i \de j_j}
\right\} \,. \label{changegamma2}
\eeqra
Deriving the above expression twice with respect to $A_\mu$ and
$A_\nu$, we get the variation in the TRG running of the self-energy
${\bf\Sigma}_{\L,\mu\nu}(q)$ induced by (\ref{gaugefix}).
Using (\ref{deltaW}), and after some algebra, we find that the first
contribution (\ref{changegamma}) can be rewritten in the form
\beq
 \frac{\i}{2} (2\pi)^8 {\rm \Tr}\left\{{\bf G_0}^{-1}_{\L,\mu\alpha}(q) 
\cdot {\bf K}_{\L, ij}(k) \cdot X_{ij\alpha\nu}(k,-k,q,-q) \right\} +
(\mu\leftrightarrow \nu)\,.
\eeq
To obtain the evolution equations for ${\bf\Sigma}_{\rm L}(q)$ and 
${\bf\Sigma}_{\rm T}(q)$, 
we have to contract with $B_{\mu\nu}(q)$ and $\half A_{\mu\nu}(q)$
respectively (see eq.~(\ref{LTSigma})).
The inverse full propagator at $q_0=0$ in a generic gauge can be written as
\beq
\left[{\bf G}_{\mu\alpha}^{-1}\right]^{11}(q_0=0, \vec{q}) = 
A_{\mu\alpha} \Delta_{\rm T}^{-1} + B_{\mu\alpha} \Delta_{\rm L}^{-1} +
D_{\mu\alpha} f_d \,.
\eeq
The $C_{\mu\alpha}$ structure is absent at $q_0=0$ due to
$T$-invariance, and moreover \mbox{${\rm Im}\,{\bf\Sigma}(q_0=0,\vec{q})=0$}.

The contribution of (\ref{changegamma}) to the
variations in the running of ${\bf\Sigma}_{\rm L}^{11}$ and 
${\bf\Sigma}_{\rm T}^{11}$ is then given by
\beqra
\ds
\L \frac{\partial\;\;\;}{\partial \L}\Delta 
{\bf\Sigma}_{\rm L}^{11}(q_0=0,\vec{q}) &=&
\i (2\pi)^8 \Delta_{\rm L}^{-1}(q) B_{\alpha\nu}(q)
\,{\rm Tr} \left\{{\bf K}_{\L,ij}
\cdot X_{ij\alpha\nu}\right\}\,, \nonumber\\
\ds
\L \frac{\partial\;\;\;}{\partial \L}\Delta 
{\bf\Sigma}_{\rm T}^{11}(q_0=0,\vec{q}) &=&
\i \frac{(2\pi)^8}{2} \Delta_{\rm T}^{-1}(q) A_{\alpha\nu}(q)
\,{\rm Tr} \left\{{\bf K}_{\L,ij}
\cdot X_{ij\alpha\nu}\right\}\,.
\eeqra
{F}rom the above expressions we see that if we compute ${\bf\Sigma}_{\rm L}$
and ${\bf\Sigma}_{\rm T}$ on the longitudinal and transverse mass shells,
respectively, the gauge dependence of (\ref{changegamma}) vanishes. 
This conclusion could be spoiled by the presence of a pole
of the function $X_{ij \alpha\nu}$ for $q$ on the same mass shell. As
discussed in \cite{KKR} the pole structure of this function is
determined by the poles of the full ghost propagator, which are 
gauge-dependent. As a consequence this coincidence between the gluon and the
ghost poles could take place only accidentally, in some particular
gauges. If only one gauge in which this is not the case exists, then
the  gauge independence of the gluon dispersion relation
can be ascertained.

The second contribution to (\ref{change}), eq.~(\ref{changegamma2}),
is absent in the
Landshoff-Rebhan formulation of covariant gauges as well as in the
TAG, so the proof is complete in these two cases.
We will then consider this contribution in the only case of practical 
interest in which this is different from zero, that is in covariant
gauges \`{a} la Hata-Kugo.
The only gauge-dependent part in the tree-level propagator is
proportional to the tensor $D^{\rho\sigma}(k)$:
\beq
\L\frac{\partial\;\;\;}{\partial \L} \Delta {\bf G}_{\L, \rho\sigma}^{-1} =
-2 i \varepsilon \frac{\partial N_\L}{\partial \L} 
\frac{\Delta \alpha}{\alpha^2} D_{\rho\sigma}(k) {\bf B}\,.
\eeq
The contribution  (\ref{changegamma2}) then gives
\beq
(2\pi)^8 \varepsilon \frac{\partial N_\L}{\partial \L} 
\frac{\Delta \alpha}{\alpha^2} {\rm Tr}\left\{D_{\rho\sigma}(k)
{\bf G}_{\L,\rho \rho^\prime}(k)\cdot {\bf B}\cdot 
{\bf G}_{\L,\sigma \sigma^\prime}(k) \cdot
\bar{\G}_{\L,\mu\nu\rho^\prime\sigma^\prime}^{aabb}(q,-q,k,-k)\right\}\,,
\label{riga2}
\eeq
where the vertex
$\bar{\G}_{\L,\mu\nu\rho^\prime\sigma^\prime}^{aabb}$ is the one 
defined in (\ref{Gse}).

Now we use the ST identity to write
\beq
- \i k_\rho {\bf G}_{\L,\rho\sigma}(k) = \alpha 
{\bf\Gamma}^{(wc)}_{\L,\sigma}(k) \cdot {\bf B}' \cdot
{\bf D}_\L(k) \,,
\eeq
where ${\bf D}_\L(k)$ is the full ghost propagator,  
${\bf\Gamma}^{(wc)}_{\L,\sigma}$ the $w-c$ vertex, 
$w_\sigma$ being the BRS-source defined in eq.~(\ref{SBRS}),
and ${\bf B}'={\rm diag}(1,-1)$.
Contracting this vertex with the four-point function in (\ref{riga2})
and using the ST identities for the three  and four  gluon vertices, 
the parenthesis in (\ref{riga2}) turns out to have the structure
\beq
\tilde{\bf G}^{-1}_{\L,\mu\alpha}(q) Z_{\alpha \nu}(q) + 
\tilde{\bf G}^{-1}_{\L,\nu\alpha}(q) Z_{\alpha \mu}(q) + 
\tilde{\bf G}^{-1}_{\L,\mu\alpha}(q)\tilde{\bf G}^{-1}_{\L,\nu\lambda}(q) 
Y_{\alpha\lambda}(q),
\eeq
where $\ds \tilde{\bf G}^{-1}_{\L,\mu\alpha}(q)\equiv 
{\bf G}^{-1}_{\L,\mu\alpha}(q)-\frac{1}{\alpha} k_\mu k_\alpha {\bf Q}[1]$.

Contracting again with $B_{\mu\nu}(q)$ and with $\half A_{\mu\nu}(q)$,
we conclude that also the contribution of (\ref{changegamma2})  
vanishes when $q$ is on the ($\L$-dependent) mass shell.

\end{document}